# *DE FACTO* CONTROL: APPLYING GAME THEORY TO THE LAW ON CORPORATE NATIONALITY

*By Russell Stanley Q. Geronimo*[*]

## INTRODUCTION

One unexamined assumption in foreign ownership regulation is the notion that majority voting rights translate to "effective control".[1] This assumption is so deeply entrenched in foreign investments law that possession of majority voting rights can determine the nationality of a corporation and its capacity to engage in partially nationalized economic activities.[2] The fact, however, is that minority stockholders can possess a degree of voting power higher than what their shareholding size might suggest.[3] Voting power is not the same as voting weight and is not measured simply by the proportion or number of votes a stockholder may cast in a stockholder meeting.[4]

An example is a voting situation requiring a simple majority (51%) with stockholders $P_1$, $P_2$ and $P_3$ having 50%, 49% and 1% voting weights, respectively.[5] While intuition tells us that $P_2$ has a disproportionate degree of control compared to $P_3$, it is not true that $P_2$ has more "effective" voting power than $P_3$.[6] And while intuition also tells us that the 1% difference in voting rights between $P_1$ and $P_2$ is insubstantial, $P_1$ still wields a more significant degree of control compared to $P_2$.[7]

*First*, note that none of the stockholders can single-handedly pass a motion, and that we have no prior knowledge of their preferences in forming alliances. Thus, it is fair to assume that each stockholder is equally likely to form a coalition with any other stockholder, which means a coalition between $P_1$ and $P_2$ is just as likely to form as a coalition between $P_2$ and $P_3$, and $P_1$ and $P_3$.[8] All the possible winning coalitions are: {$P_1$, $P_2$}, {$P_1$, $P_3$} and {$P_1$, $P_2$, $P_3$}. *Second*, note that $P_1$ only needs one more vote to muster the required minimum votes to pass a desired stockholder resolution, and it is a matter of indifference to $P_1$ whether that vote comes from $P_2$ or $P_3$. *Third*, $P_2$ cannot prevail if he forms an alliance with $P_3$ alone. $P_1$ is a swing voter in all these three instances, while $P_2$ is a swing voter in only one instance. By "swing voter", we mean a voter who can make

---

[*] Juris Doctor, University of the Philippines – College of Law

[1] *Gamboa vs. Teves*, G.R. No. 176579, October 09, 2012
[2] The foreign equity limitation "must [likewise] apply separately to each class of shares, whether common, preferred non-voting, preferred voting or any other class of shares." (*Gamboa vs. Teves*, G.R. No. 176579, October 9, 2012)
[3] Leech, D., *Shareholder Voting Power and Corporate Governance: A Study of Large British Companies*, 27 NORDIC JOURNAL OF POLITICAL ECONOMY 1, 33-54 (2001)
[4] Holler, M.J., *Forming Coalitions and Measuring Voting Power*, 30 POLITICAL STUDIES 2, 262-271 (1982)
[5] A classic example in introductions to weighted voting systems. *See* Citroen, R., and Penn, M., *Fair Division in Theory and Practice*, available at: http://www.cs.wustl.edu/~cytron/fdiv/PDFs/8.pdf (accessed on 16 August 2016)
[6] Prigge, S. and Kehren, S., *Ownership Structure Metrics*, INTERNATIONAL CORPORATE GOVERNANCE AFTER SARBANES-OXLEY, Paul Ali, Greg N. Gregoriou, eds., Wiley (2006)
[7] *Id.*
[8] Applying the principle of *a priori* probability in weighted voting systems. *See* Stenlund, H., and Lane, J, *The Structure of Voting-Power Indices*, 18 QUALITY AND QUANTITY 367-375 (1984)

a coalition lose by dropping out of the coalition.[9] $P_2$ cannot block a motion by $P_1$ once the latter forms an alliance with $P_3$. $P_2$ also cannot make the grand coalition of all three stockholders lose by dropping out. In this context, $P_2$ is in the same position as $P_3$.

There is no doubt, therefore, that $P_1$ has a voting power disproportionately larger than $P_2$, and we are only talking about a difference of 1% in voting rights between them. There is also no doubt that $P_2$ has a voting power equivalent to that of $P_3$, even though they have a seemingly substantial difference of 48% voting rights. This simple voting situation demonstrates that voting weight has a non-linear relationship with actual voting power.[10] Clearly, "the largest shareholders are not always winners, nor are the smaller shareholders predestined losers."[11]

A share of stock represents a bundle of stockholder rights,[12] which include economic rights and control rights.[13] Economic rights pertain to pecuniary interests, such as the right to dividends, right to sell shares, and right to a portion of the net asset value upon dissolution and liquidation of the company.[14] Control rights, on the other hand, allow stockholders to participate in making business decisions.[15] These are expressed in terms of voting rights in stockholder meetings, where one share of stock is usually equal to one vote.[16] Increasing the number of shares (i.e. the shareholding size) leads to an increase in economic and control rights. In the case of economic rights, the increase is linear and positively correlated.[17] Hence, in a corporation declaring dividends, having 10% shareholding size entitles the stockholder to 10% of the total dividends declared, 20% shareholding size entitles him to 20%, and so on.

This notion, however, may erroneously lead us to assume that the relationship between shareholding size and control rights is also linear and positively correlated. In fact, when we increase shareholding size, control rights do not increase in the same manner as economic rights. This assaults our basic intuition about the nature of control rights because we know that a higher number of shares results in a higher voting weight.

---

[9] German, A., Katz, J., and Tuerlinckx, F., *The Mathematics and Statistics of Voting Power*, 17 STATISTICAL SCIENCE 4, 420-435 (2002)

[10] For a discussion on the formal properties of voting power measurements, *see* Holler, M.J., and Napel, S., *Monotonicity of Power and Power Measures*, 56 THEORY AND DECISION 1-2, 93-111 (2004)

[11] Bajuri, N., Chakravarty, S., and Hashim, N., *Analysis of Corporate Control: Can the Voting Power Index Outshine Shareholding Size?*, 10 ASIAN ACADEMY OF MANAGEMENT JOURNAL OF ACCOUNTING AND FINANCE 1, 75-94 (2014)

[12] Micheler, Eva, *Custody Chains and Remoteness - Disconnecting Investors from Issuers* (2014), available at: http://ssrn.com/abstract=2413025 ("Securities are a bundle of rights that investor have against issuers.")

[13] Hu, Henry T. C. and Black, Bernard S., *Debt, Equity and Hybrid Decoupling: Governance and Systemic Risk Implications*, 14 EUROPEAN FINANCIAL MANAGEMENT 663 (2008)

[14] Economic rights are also called "cash flow rights". *See* Ranade, S.M., *Separation of Voting Rights from Cash-Flow Rights in Corporate Law: In Search of the Optimal*, Warwick School of Law Research Paper (2013), available at SSRN: http://ssrn.com/abstract=2246757

[15] Control rights are also called "voting rights", since it is through the exercise of formal voting power that stockholders can pass shareholder resolutions. *See* Dong, L. and Uchida, K. and Hou, X., *How Do Corporate Control Rights Transactions Create Shareholder Value? Evidence from China* (2014), available at SSRN: http://ssrn.com/abstract=2396514

[16] *Id.*

[17] *Supra* note 10.

Based on the simple voting situation we have shown, we see that examining and comparing the voting weights of stockholders does not give a true description of their voting power. This is because *voting weight is not equivalent to voting power*.[18] The 1% voting weight difference between $P_1$ and $P_2$ makes a true difference in determining the outcome of the stockholder meeting in a way that the 48% difference between $P_2$ and $P_3$ did not. An increase in 1% shareholding size may result in an equivalent increase of 1% voting weight, but it does not necessarily result in an increase of 1% voting power.

This leads to anomalous situations where foreign minority stockholders have *de facto* control of a Filipino corporation engaged in a partially nationalized economic activity, effectively subverting the nationalist policy of the 1987 Constitution and *Gamboa vs. Teves* on foreign equity limitations.[19] In the example, assume that $P_3$ is a foreigner and the Filipino corporation is engaged in an industry with 20% foreign equity limitation. While $P_3$'s voting weight of 1% falls comfortably below the equity cap, $P_3$ has a *de facto* or effective control of 25%, equal to $P_2$.[20] In the succeeding sections, we shall propose and demonstrate a method for calculating "effective control" based on given voting thresholds and voting weights. We shall also show instances where the "effective control" of a foreign minority stockholder appears to comply with foreign equity caps, but has a "real" voting power grossly beyond the allowable threshold.

This problem exists because the Control Test equates voting power with voting weight,[21] when the fact is that voting weight can be higher than or less than the actual voting power of stockholders.[22] By overstating or understating voting power, the Control Test permits situations where a Philippine national is actually controlled by foreign stockholders, or a foreign national is effectively controlled by Filipino stockholders. By relying solely on the "absolute"[23] voting weight of one stockholder, the Control Test fails to consider a host of factors that determine the true nature of voting power, namely:

(1) Number of stockholders;
(2) Minimum votes required to pass a stockholder resolution;
(3) Amount of voting rights held by one stockholder in relation to other stockholders;
(4) Possibility of forming a coalition of stockholders;
(5) Number of times that a stockholder can be a swing voter; and
(6) The size of the public float, if any.[24]

**FOREIGN CONTROL OF STRATEGIC INDUSTRIES AS A GEOPOLITICAL RISK**

Why should *de facto* foreign control of sensitive economic activities be a concern for the Philippines? Developed nations like Australia, Canada, United Kingdom and United States operate under a system of free trade, where foreign ownership limitations and other citizenship purity protocols in economic activities are considered sources of market distortions and inefficiencies.[25] For these countries, it is a matter of indifference whether corporations operating vital industries are foreign- or domestic-owned.

---

[18] Leech, D., and Manjón, M.C., *Corporate Governance and Game Theoretic Analyses of Shareholder Power: The Case of Spain*, 35 APPLIED ECONOMICS 7, 847-858 (2003)
[19] *See* Executive Order No. 858 dated 05 February 2010 for foreign equity restrictions in various industries
[20] We derived this figure using the Banzhaf Voting Power Index. *See* Straffin, P.D., *The Shapley-Shubik and Banzhaf Power Indices as Probabilities*, THE SHAPLEY VALUE: ESSAYS IN HONOR OF LLOYD S. SHAPLEY (1988)
[21] *Gamboa vs. Teves*, G.R. No. 176579, October 09, 2012
[22] *Supra* note 20.
[23] As opposed to "relative" voting weight. The absolute voting weight looks at the shareholding size of one stockholder, while relative voting weight looks at the distribution of voting weights among all stockholders.
[24] *Supra* note 6.

There is strong political pressure in the Philippines to relax its laws on foreign investment limitations. In August 2016, newly elected President Rodrigo Duterte expressed willingness for a constitutional amendment to ease foreign ownership restrictions imposed by the 1987 Constitution in land ownership, in the exploitation, development and utilization of natural resources, and in the operation of public utilities.[26]

The reality, however, is that foreign control of sensitive economic activities is a major source of geopolitical risk. Even developed nations operating under a free trade are now beginning to recognize this. "Increasingly, corporations are political tools used by nations to exert influence over other nations. In times of peace and economic prosperity, foreign control of strategic industries and infrastructure may not be an immediate threat. But during major economic recessions — or, worse, times of geopolitical upheaval and war — the loss of ownership and full control of national industries can be catastrophic."[27]

One example is the Russia-Ukraine gas dispute in 2006. Gazprom, a Russian-owned gas company, wanted to increase the price of oil passing through Ukraine from USD 50 to USD 230 per 1,000 cubic metres. Ukraine rejected the offer. In response, Gazprom cut off the supply of gas to Ukraine, causing shortage of gas supply in the whole European Union. Many believed that it was not a purely commercial dispute, and that it was an instance where a "foreign company's decisions become an extension of the [foreign] government's policy decisions rather than the company's commercial interests."[28]

In the same year, there was a national security debate in the U.S. concerning the attempted foreign takeover of six major seaports by Dubai Ports World, a government-owned corporation based in the United Arab Emirates (UAE). Many national security analysts believed that this would render the U.S. susceptible to terrorist attacks, considering the large number of containers entering the U.S. and the possibility of importing illegal weapons and international transport of terrorists. This led to the passage of the Foreign Investment and National Security Act of 2007, which strengthens the power of the U.S. government to review foreign investments in strategic industries.[29]

---

[25] *Airbus and the Perils of Foreign Ownership*, theTrumpet.com, October 26, 2006, available at: https://www.thetrumpet.com/article/2969.2.0.0/world/globalization/airbus-and-the-perils-of-foreign-ownership

[26] France-Presse, A., Duterte wants to open Philippines to foreign investors: aide, May 13, 2016, ABS-CBN News, available at: http://news.abs-cbn.com/business/05/12/16/duterte-wants-to-open-philippines-to-foreign-investors-aide

[27] *Supra* note Airbus

[28] Masters, J., Foreign Investment and U.S. National Security, Council on Foreign Relations, September 27, 2013, available at: http://www.cfr.org/foreign-direct-investment/foreign-investment-us-national-security/p31477

[29] Mostaghel, D.M., *Dubai Ports World under Exon-Florio: A Threat to National Security or a Tempest in a Seaport*, 70 ALB. L. REV. 583 (2006-2007)

Another example is the rise of Rosatom, a Russian-owned nuclear corporation, which operates in 40 countries and has 29 ongoing global projects, including countries like Turkey, Armenia, Finland, Belarus, Vietnam, Bangladesh, India and China. Many believe that Rosatom is pivotal in Russia's nuclear diplomacy.[30]

As of October 2016, sovereign wealth funds (SWFs), which are government-owned foreign investment vehicles, are operating with USD 7.39 trillion assets all over the world. This raises several national security concerns for host countries receiving their investments, including the "destabilization of the financial markets (to the detriment of the host country), protection of SWF home-country industries at the expense of the host country's industries, and the expropriation of technology[.]"[31] One of the criticisms against SWFs is that most of them are based in authoritarian regimes facing risks of political instability, and that these funds could be utilized to further their international political agenda.

The Philippines is in the midst of a geopolitical game involving China, U.S., and Russia, and it is not far-fetched to imagine that foreign investments will play a crucial role in the brinkmanship of world superpowers in their struggle to protect their maritime interests. With the aim of pursuing an independent foreign policy, President Duterte announced opening economic alliances with China and Russia, including the development of vital infrastructure projects, like railways and seaports. This makes the analysis of *de facto* control of corporations all the more urgent.

## THE STOCKHOLDER MEETING AS A WEIGHTED VOTING GAME

We can remedy the limitations of the Control Test by adopting multiple-factor voting power measurements, such as the Banzhaf Voting Power Index in the field of cooperative game theory. We shall begin by modeling a traditional stockholder meeting as a weighted voting game.

There are two voting systems in Philippine corporation law: the one person-one vote system and the one share-one vote system.[32] In the former system, all voters have equal voting power.[33] This is the default situation in board meetings, where each board member present is entitled to only one vote, regardless of whether he is a nominee of a stockholder having disproportionate ownership interest in the corporation.[34] The same default rule applies in non-stock corporations, where each member is entitled to only one vote unless otherwise provided in the by-laws.[35]

---

[30] Dobrev, B., *Rosatom & Russia's Nuclear Diplomacy*, Geopolitical Monitor, May 17, 2016, available at: https://www.geopoliticalmonitor.com/rosatom-russias-nuclear-diplomacy/

[31] Hemphill, T.A., Sovereign Wealth Funds: National Security Risks in a Global Free Trade Environment, 51 Thunderbird International Business Review 6 (2009)

[32] *See* various voting rules in B.P. No. 68 (*The Corporation Code of the Philippines*).

[33] Hayden, G.M., *The False Promise of One Person, One Vote*, 102 MICHIGAN LAW REVIEW 2, 213-267 (2003)

[34] This is without prejudice to the power of the corporation to adopt by-laws prescribing the manner of voting. *See* Sec. 46, B.P. No. 68.

[35] Sec. 89, B.P. No. 68 ("Unless so limited, broadened or denied, each member, regardless of class, shall be entitled to one vote.")

In the one share-one vote system, a voter can have higher or lesser voting power compared to others, depending on the amount of voting shares held.[36] This is the rule in stockholder meetings of stock corporations, where different percentage holdings yield different amount of votes per stockholder.[37] This is also the rule where fundamental matters require the participation of preferred shareholders.[38]

In the one person-one vote system, only two elements are important in determining the results of a voting situation: the number of voters and the minimum number of votes required to pass a resolution.[39] In the one share-one vote system, one additional element is essential: the number of votes that each voter is entitled to cast.[40]

These three variables qualify the one share-one vote system in stockholder meetings as a weighted voting game: the *players*, the *quota*, and the *weight*.[41] The *players* represent the stockholders entitled to vote.[42] The *quota* denotes the minimum number of votes required to pass a stockholder resolution.[43] It is otherwise called the decision threshold, which may be majority (51%), super-majority (67%), unanimous (100%), or any other threshold specified in the by-laws.[44] The *weight* is the number of votes that each player is entitled to cast. It is otherwise called the shareholding size.[45] A *game* represents a voting situation involving only two alternative motions: "yes" and "no", where "abstain" is counted as "no".[46]

The one person-one vote system can evolve into a weighted voting game, and a weighted voting game can evolve into a one person-one vote system. Consider the following scenarios:

1. *When one person-one vote system becomes a weighted voting game.* – In a one person-one vote system, the concept of weight is immaterial if viewed from the perspective of individual voters. Voting power is represented as $1/N$, where $N$ is the total number of players. Thus, if $N = 10$, the voting power of $P_1$ is 10%, which is the same for all other players. The concept of weight becomes material only if a group of voters is conceived as a coalition, in

---

which case the one person-one vote system also becomes a weighted voting game from the perspective of the coalition of voters.[47] We consider each coalition as a single player, and the weight is the number of voters in a coalition.[48] Thus, in a board of directors composed of six Filipinos and four foreigners, the Filipino coalition has 60% weight and the foreign coalition has 40% weight.[49] Here we have a situation where a board meeting, which is a one person-one vote system, being reconfigured as a weighted voting game.

2. *When a weighted voting game becomes a one person-one vote system.* – If all players are required to have one vote to pass a stockholder resolution, then the weights become immaterial, just like in a one person-one vote system.[50] This is the case where no individual player or coalition of players can muster enough votes to meet the quota, except the grand coalition of all players. In short, the voting situation *de facto* requires a unanimous vote. Hence, given stockholders $P_1$, $P_2$ and $P_3$ with respective weights of 60%, 20% and 20%, and a quota of 81% votes, $P_1$ will always require the votes of $P_2$ and $P_3$ to pass a stockholder resolution. The {60%, 20%, 20%} voting weight distribution is irrelevant because even though two stockholders form a coalition, they cannot muster 81% of the requisite votes. It is as though the voting power of each player is 1/3, or more generally, $1/N$, which is precisely the voting power formula in a one person-one vote system.

In modeling the stockholder meeting as a weighted voting game, the absolute voting weight of one stockholder is not a sufficient indicator of his voting power.[51] In order to accurately describe the stockholder's voting power, it is necessary to consider how all the pertinent variables of a weighted voting game (the number of players, the quota, the weight, and the coalitions) relate to one another.[52] To facilitate the discussion, we shall adopt the formal notation of a weighted voting game to represent a stockholder meeting, as follows:

$$\{q: w_1, w_2 \ldots w_N\}$$

In this notation, $q$ represents the quota; $w_1, w_2 \ldots w_N$ represents the individual stockholders with their respective voting weights; and $N$ is the total number of stockholders. Hence, in a stockholder meeting requiring a simple majority or 51% to pass a resolution, with five stockholders having a percentage holding distribution of 50%, 25%, 10%, 10% and 5%, the voting game is expressed as {51: 50, 25, 10, 10, 5}. In a stockholder meeting requiring 2/3 or 67% votes to pass a resolution, given the same players and weights, the voting game is expressed as {67: 50, 25, 10, 10, 5}.[53]

---

[47] Taylor, A., and Zwicker, W., *A Characterization of Weighted Voting*, 115 PROCEEDINGS OF THE AMERICAN MATHEMATICAL SOCIETY 4, 1089-1094 (1992)
[48] Lucas, W.F., *Measuring Power in Weighted Voting Systems*, POLITICAL AND RELATED MODELS, Springer New York, 183-238 (1983)
[49] Accordingly, we can also apply the concept of "voting power" as discussed in this article whenever we conceive the board of directors as a coalition of Filipino and foreign directors.
[50] Feingold, R.D., *Representative Democracy versus Corporate Democracy: How Soft Money Erodes the Principle of One Person, One Vote*, 35 HARVARD JOURNAL ON LEGISLATION 377 (1998)
[51] Chen, X. and Sinha, A.K., *Two Proxies for Shareholder Influence: A Case of Payout Policy* (2009), available at: http://ssrn.com/abstract=1522504
[52] Leech, D., *An Empirical Comparison of the Performance of Classical Power Indices*, 50 POLITICAL STUDIES 1, 1-22 (2002)

## THE STOCKHOLDERS AS PLAYERS

For every stockholder meeting, there are three possible types of voting stockholders: a "dictator", a "dummy", and a player with veto power.[54] A dictator has the power to pass a resolution single-handedly.[55] A dummy is one whose voting power is immaterial in passing a resolution.[56] And a player with veto power is one whose vote is indispensable to pass a resolution, but cannot pass a resolution single-handedly.[57]

The dictator status represents the highest degree of control possible in a given voting situation. The dummy represents the lowest possible degree of control. And veto power represents joint or equal control shared between or among stockholders.[58] The commonality underlying these three types of stockholders is that their respective degrees of control are not solely determined by voting weight.[59] This demonstrates the notion that voting weight alone is not the sole factor of voting power.[60]

We shall examine each of these stockholder types in the succeeding sections. We shall also demonstrate the inadequacy of voting weight in determining dictator or dummy status and the presence of veto power.

## DICTATOR STOCKHOLDERS

A stockholder with a sufficiently large voting weight to pass a resolution single-handedly renders the voting weight and voting power of other stockholders immaterial.[61] This "dictator" status satisfies the following condition in a stockholder meeting:

$$w_i \geq q$$

The voting weight ($w_i$) of a stockholder must be equal to or higher than the quota ($q$).[62] This suggests that voting weight alone is insufficient information to conclude that a

---

stockholder has dictator status. The decision threshold, which may be a simple majority (51%), super-majority (2/3 or 67%) or unanimous vote (100%), is a critical element. Consider the following voting situations in a stockholder meeting:

1. {51: 51, 49}
2. {67: 51, 49}
3. {100: 99, 1}

In the first example, the quota is 51%, with stockholders $P_1$ and $P_2$ having 51% and 49% voting rights, respectively. Since $P_1$ can single-handedly pass a resolution, he has a dictator status in a voting game. The situation is effectively the same as a voting situation with {100: 100, 0} voting power distribution. $P_1$ has an effective voting power of 100% because he does not need the cooperation of $P_2$ to muster enough votes in a stockholder meeting. And while the 49% voting rights of $P_2$ may appear to be a considerable amount of voting power, $P_2$ has an effective voting power of only 0% because his vote will never be relevant in determining the outcome of the stockholder meeting. In short, it does not matter whether $P_2$ has 0% or 49% voting rights, or any arbitrary shareholding size between 0% and 49%, as long as the voting weight of $P_1$ is equal to or greater than the quota of 51%.

In the second example, the quota is a super-majority requirement of 2/3 votes or 67%, with stockholders $P_1$ and $P_2$ having the same voting rights as in the first example. The only difference between the first and second examples is the quota. However, this difference makes $P_1$ lose his dictator status. In fact, $P_1$ and $P_2$ have joint control in the corporation, with an effective voting power distribution of {100: 50, 50}. It is a matter of indifference whether $P_1$'s voting weight of 51% is higher than $P_2$'s voting weight of 49%. The voting rights differential of 2% is irrelevant in determining the outcome of the stockholder meeting.

In the third example, the quota requires a unanimous vote, with stockholder $P_1$ having 99% voting rights and $P_2$ having 1% voting rights. Their percentage holdings differ by a wide margin. However, considering a quota of 100%, their effective voting power distribution is {100: 50, 50}, which is the same as the effective voting power distribution in the second example. In the second example, the difference in voting rights is 2%. In the third example, the difference is 98%. These differences, however, are immaterial in determining the final outcome of the stockholder meeting. $P_2$'s 1% voting weight is indispensable to pass a stockholder resolution.

The differences in voting weight distribution and effective voting power distribution in the three scenarios are summarized as follows:

| Voting Weight Distribution | Voting Power Distribution |
|---|---|
| {51: 51, 49} | {100: 100, 0} |
| {67: 51, 49} | {100: 50, 50} |
| {100: 99, 1} | {100: 50, 50} |

The left column describes the *de jure* allocation of control in the corporation, which uses "voting weight" as a criterion, while the right column describes the *de facto* allocation of control, which uses the concept "voting power".[63]

In these examples, only the first has a dictator. This shows that a dictator status, which represents the highest degree of control possible in a stockholder meeting, is a relationship between two factors: the quota and the voting weight of a stockholder in relation to the voting weight of the other stockholder/s. This shows further that merely relying on the absolute voting weight of one stockholder gives incomplete information about his true voting power.[64]

**DUMMY STOCKHOLDERS**

A stockholder whose voting weight is immaterial in determining the outcome of a stockholder meeting is a "dummy", which represents the lowest degree of control possible in a corporation.[65] A stockholder is a dummy if the following conditions are satisfied:

1. There is no single instance that he can make any possible coalition of stockholders prevail in a stockholder meeting by joining; and
2. There is no single instance that he can make any coalition lose by dropping out.

Whenever there is a dictator, all other stockholders are dummies.[66] This is true in the first example in the previous section, involving the voting rights distribution {51: 51, 49}, with $P_1$ as dictator. $P_2$ can neither help $P_1$ prevail nor block his motion in a stockholder meeting because $P_1$'s voting weight is already equal to the quota.

It is also possible to have dummy stockholders where there is no dictator. Moreover, a stockholder can be a dummy even though he has a nearly equal voting weight as the other stockholders. Consider the following illustrations:

1. {51: 49.5, 49.5, 1.0}
2. {51: 50, 49, 1}
3. {67: 34, 34, 32}

The first example has a quota of 51% and stockholders $P_1$, $P_2$ and $P_3$ have voting weights of 49.5%, 49.5% and 1.0%, respectively. $P_3$ is a dummy because there is no single instance that he can make a coalition with $P_1$ or $P_2$ prevail in a voting situation. Furthermore, there is no single instance that he can make the grand coalition of all stockholders lose in a voting situation by dropping out. In short, $P_3$'s voting weight is immaterial in determining the outcome of a stockholder meeting.

Compare this with the second example, which has the same quota as the first example, but with a very miniscule modification in the voting rights of $P_1$ and $P_2$, with respective voting weights of 50% and 49%. Here, we merely shifted .5% from $P_2$ to $P_1$'s voting weight, while $P_3$'s voting weight of 1% remains unchanged from the first example.

---

[63] *Supra* note 18 (distinguishing voting weight and voting power)
[64] Lucas, W.F., *Measuring Power in Weighted Voting Systems*, POLITICAL AND RELATED MODELS, Springer New York 183-238 (1983)
[65] *Supra* note 61.
[66] *Id.*

Notice that this minor change of .5% in the voting rights of *other* stockholders made $P_3$ lose his dummy status. Suddenly, $P_3$ becomes a critical voter and can make $P_1$ win or lose without the cooperation of $P_2$. $P_3$'s voting weight of 1% may be disproportionately lower than $P_2$'s voting weight of 49%, but $P_3$'s voting power is effectively or *de facto* equal to $P_2$.

The third example shows that a nearly equal voting weight can still result in disproportionate degrees of voting power. With a quota of 2/3 or 67% super-majority votes, and a voting rights distribution of 34-34-32, $P_3$ has a nearly equal voting weight as $P_1$ and $P_2$. $P_3$'s voting weight differs only by a margin of 2%, yet it is inaccurate to say that $P_3$ has equal voting power as $P_1$ and $P_2$. The truth is that only $P_1$ and $P_2$ have effective control of the corporation, with *de facto* control of 50-50 voting power. Meanwhile, $P_3$ has 0% voting power. This is because $P_1$ or $P_2$ cannot muster enough votes to pass a resolution by forming a coalition with $P_3$. Second, a coalition composed of $P_1$ and $P_2$ is the only possible winning coalition. Third, in a grand coalition composed of all stockholders, $P_3$ is not a critical voter—i.e., dropping out will not make the coalition lose. This renders $P_3$ a dummy.

The differences in voting weight distribution and effective voting power distribution in the two scenarios are summarized as follows:

| Voting Weight Distribution | Voting Power Distribution |
|---|---|
| {51: 49.5, 49.5, 1.0} | {100: 50, 50, 0} |
| {51: 50, 49, 1} | {100: 50, 25, 25} |
| {67: 34, 34, 32} | {100: 50, 50, 0} |

Again, the left column describes the *de jure* allocation of control in the corporation, while the right column describes the *de facto* allocation of control. Only the first and third examples have dummies, represented by 0% voting power.[67]

This demonstrates the weakness of the Control Test in describing the true voting power of stockholders. *First*, the voting weights of other stockholders can modify the voting power of a stockholder, even though the latter's voting weight remains unchanged.[68] *Second*, we cannot judge the voting power of a stockholder merely by looking at the magnitude of his voting weight. A less than 1% shift in voting weight, or a voting rights differential of 2%, can modify the total voting power distribution in the whole corporation.[69] *Third*, a stockholder can have as many votes as other stockholders and yet still be a dummy.[70]

VETO POWER

---

[67] *Supra* note 18.
[68] Poulsen, T., Strand, T., and Thomsen, S., *Voting Power and Shareholder Activism: A Study of Swedish Shareholder Meetings*, 18 CORPORATE GOVERNANCE: AN INTERNATIONAL REVIEW 4, 329-343 (2010)
[69] Leech, D., *Ownership Concentration and the Theory of the Firm: A Simple-Game-Theoretic Approach*, THE JOURNAL OF INDUSTRIAL ECONOMICS 225-240 (1987)
[70] Crama, Y., and Leruth, L., *Power Indices and the Measurement of Control in Corporate Structures*, 15 INTERNATIONAL GAME THEORY REVIEW 3 (2013)

Veto power is that degree of voting power that can block a motion, but cannot on its own pass a motion.[71] It is a lower degree of control compared to a dictator status. A stockholder with veto power satisfies the following two conditions:

$$w_1 < q$$

$$\left(\sum w_i\right) - w_1 < q$$

The first condition is that the stockholder's voting weight ($w_1$) should be less than the quota ($q$); otherwise, he is a dictator. The second condition is that the total voting weights of all stockholders ($\sum w_i$), minus the stockholder's voting weight ($w_1$), should be less than the quota ($q$). This means that even if all other stockholders form a coalition, they cannot muster the required minimum votes to pass a stockholder resolution.[72] The stockholder's vote is indispensable, but he himself cannot pass a resolution single-handedly. He has power to prevent a motion from passing, but he has no unilateral power to pass a motion. He can make the coalition of all other stockholders win or lose in a stockholder meeting.[73]

In a corporation with only two stockholders, a stockholder with veto power has a *de facto* control of 50% voting power, regardless of what his shareholding size or voting weight might be.[74] These two conditions that create veto power describe a relationship between voting weight distributions and the quota. As in the previous sections, we cannot deduce whether a stockholder has veto power based on his absolute voting weight alone.[75] Consider the following illustrations:

1. {51: 50, 25, 25}
2. {67: 40, 30, 30}
3. {100: 33, 33, 33, 1}

In the first example, $P_1$ cannot pass a motion single-handedly because his voting weight of 50% is less than the quota of 51%. The combined voting weight of $P_2$ and $P_3$, which is 50%, is also less than the quota. $P_1$'s vote is indispensable if $P_2$ and $P_3$ want to pass a motion. He can likewise make the coalition of $P_2$ and $P_3$ lose in the stockholder meeting. The second example has essentially the same voting power setup as in the first example, with $P_1$ having a veto power because $P_2$ and $P_3$ absolutely require his cooperation to pass a motion.

The third example exemplifies the non-monotonicity between voting weight and voting power. Since the quota requires a unanimous vote, $P_4$'s measly voting weight of 1% is, in reality, equivalent to 25% voting power. This is also an instance where a weighted voting game like a stockholder's meeting evolves into a one person-one vote system, where each voter has *de facto* equal voting power, calculated simply as $1/N$, where $N$ signifies the number of voting stockholders.

---

[71] Newman, D.P., *The SEC's Influence on Accounting Standards: The Power of the Veto*, JOURNAL OF ACCOUNTING RESEARCH 134-156 (1981)
[72] *Supra* note 61.
[73] *Id.*
[74] Straffin, P.D., *Homogeneity, Independence, and Power Indices*, 30 PUBLIC CHOICE 1, 107-118 (1977)
[75] Leech, D., and Manjón, M.C., *Corporate Governance in Spain (With an Application of the Power Indices Approach)*, 13 EUROPEAN JOURNAL OF LAW AND ECONOMICS 2, 157-173 (2002)

The differences in voting weight distribution and effective voting power distribution in the three scenarios are summarized as follows:

| Voting Weight Distribution | Voting Power Distribution |
|---|---|
| {51: 50, 25, 25} | {100: 50, 25, 25} |
| {67: 40, 30, 30} | {100: 50, 25, 25} |
| {100: 33, 33, 33, 1} | {100: 25, 25, 25, 25} |

Based on the previous two sections and this section, a dictator stockholder has 100% voting power, a dummy stockholder has 0% voting power, while a stockholder with veto power has 50% voting power or x% voting power equal to all other stockholders. These figures reflect *de facto* or effective control regardless of the magnitude of their voting weights.

**STOCKHOLDER COALITIONS**

The Control Test fails to consider the possibility of stockholder coalitions, or situations where a stockholder will join other stockholders to pass a motion through their combined voting weight.[76] The reality is that a given voting weight can have varying degrees of voting power depending on whether it is sufficiently relevant to make alliances win or lose in a stockholder meeting.[77] A stockholder's voting weight of x% is of value to another stockholder if their combined voting weights can pass a resolution, and is of less value if it cannot.[78] To facilitate discussion, we shall adopt the following notations to denote a stockholder coalition:

$$\{P_1, P_2, P_3\}$$

A coalition composed of all stockholders is called the "grand coalition". A coalition that can muster sufficient votes to meet the quota is called the "winning coalition". A coalition that has insufficient votes to meet the quota is a "losing coalition". The combined voting weight of stockholders in a coalition is called the "coalition weight". The coalition weight of a winning coalition is always equal to or higher than the quota, and the coalition weight of a losing coalition is always lower than the quota.[79]

How do stockholder coalitions affect the individual voting power of a stockholder? A stockholder who can make a coalition win or lose has higher voting power compared to a stockholder whose voting weight is irrelevant to a coalition. In short, a stockholder who is a "swing voter" has more degree of control. To be a swing voter, the voting weight of a stockholder can either turn a losing coalition into a winning coalition or a winning coalition into a losing coalition. A stockholder who cannot make a losing coalition win by joining, or a winning coalition lose by dropping out, is not a swing voter and has low degree of control. We shall discuss this more thoroughly in the section on Critical Stockholders.[80]

---

[76] Kulpa, A.M., *The Wolf in Shareholder's Clothing: Hedge Fund Use of Cooperative Game Theory and Voting Structures to Exploit Corporate Control and Governance*, 6 UC DAVIS BUS. LJ 78-183 (2005).

[77] Crama, Y., et al., *Corporate Governance Structures, Control and Performance in European Markets: A Tale of Two Systems*, No. CORE Discussion Papers (1999/42), UCL (1999)

[78] Crama, Yves, et al., *Corporate Control Concentration Measurement and Firm Performance*, 17 SOCIAL RESPONSIBILITY: CORPORATE GOVERNANCE ISSUES, RESEARCH IN INTERNATIONAL BUSINESS AND FINANCE (2003)

[79] *Supra* note 58.

How do we know which stockholder coalitions will form? The answer is that we can never know just by looking at an arbitrary list of stockholders and their voting weights. Since we have no knowledge of the preferences of stockholders in forming alliances, it is necessary to list *all* possible coalitions for every given set of stockholders.[81] The total possible stockholder coalitions can be obtained through the following[82]:

$$2^N - 1$$

*N* denotes the total number of stockholders. The formula counts a lone stockholder as a single coalition. Hence, in a corporation with two stockholders, there are 3 possible coalitions: $\{P_1\}$, $\{P_2\}$, and $\{P_1, P_2\}$. In a corporation with three stockholders, there are 7 possible coalitions: $\{P_1\}$, $\{P_2\}$, $\{P_3\}$, $\{P_1, P_2\}$, $\{P_2, P_3\}$, $\{P_1, P_3\}$, and $\{P_1, P_2, P_3\}$. Consider the following voting situations:

1. {51: 50, 49, 1}
2. {67: 40, 30, 30}

In the first example, the total possible coalitions and the voting outcome for each coalition are illustrated as follows:

| Possible Coalitions | Coalition Weight | Voting Outcome |
|---|---|---|
| $\{P_1\}$ | 50 | Losing Coalition |
| $\{P_2\}$ | 49 | Losing Coalition |
| $\{P_3\}$ | 1 | Losing Coalition |
| $\{P_1, P_2\}$ | 99 | Winning Coalition |
| $\{P_2, P_3\}$ | 50 | Losing Coalition |
| $\{P_1, P_3\}$ | 51 | Winning Coalition |
| $\{P_1, P_2, P_3\}$ | 100 | Winning Coalition |

In the second example, the total possible coalitions and the voting outcome for each coalition are illustrated as follows:

| Possible Coalitions | Coalition Weight | Voting Outcome |
|---|---|---|
| $\{P_1\}$ | 40 | Losing Coalition |
| $\{P_2\}$ | 30 | Losing Coalition |
| $\{P_3\}$ | 30 | Losing Coalition |

---

[80] *Supra* note 6.
[81] *Id.*
[82] *Id.*

| | | |
|---|---|---|
| $\{P_1, P_2\}$ | 70 | Winning Coalition |
| $\{P_2, P_3\}$ | 60 | Losing Coalition |
| $\{P_1, P_3\}$ | 70 | Winning Coalition |
| $\{P_1, P_2, P_3\}$ | 100 | Winning Coalition |

## CRITICAL STOCKHOLDERS

A critical stockholder is a swing voter in a stockholder coalition.[83] He can make a winning coalition lose by dropping out, or he can make a losing coalition win by joining.[84] Therefore, a critical stockholder satisfies the following condition:

$$w_c - w_i < q$$

In this condition, $w_c$ represents the coalition weight; $w_i$ represents the voting weight of a stockholder who is a member of the coalition; and $q$ represents the quota.[85] We measure voting power by the number of times that the stockholders are critical stockholders, given all possible stockholder coalitions. Consider the following voting situations:

1. {51: 50, 49, 1}
2. {67: 50, 49, 1}
3. {51: 40, 30, 30}
4. {67: 40, 30, 30}

For the first example, $P_1$ has the highest voting power while $P_2$ and $P_3$ have equal voting powers. Our basis for this conclusion is that, given all 7 possible stockholder coalitions, $P_1$ is a critical stockholder in three instances, while $P_2$ and $P_3$ are critical stockholders once. This is illustrated as follows:

| Possible Coalitions | Coalition Weight | Voting Outcome | Critical Stockholders | | |
|---|---|---|---|---|---|
| | | | $P_1$ | $P_2$ | $P_3$ |
| $\{P_1\}$ | 50 | Losing Coalition | | | |
| $\{P_2\}$ | 49 | Losing Coalition | | | |
| $\{P_3\}$ | 1 | Losing Coalition | | | |
| $\{P_1, P_2\}$ | 99 | Winning Coalition | | | |
| $\{P_2, P_3\}$ | 50 | Losing Coalition | | | |
| $\{P_1, P_3\}$ | 51 | Winning Coalition | | | |
| $\{P_1, P_2, P_3\}$ | 100 | Winning Coalition | | | |
| Number of Times that Stockholder is Critical | | | 3 | 1 | 1 |

In the second example, we have the same voting rights distribution as in the first example, but we changed the quota from a simple majority of 51% to a super-majority of 67%. This also modifies the voting power of the stockholders, with $P_1$ and $P_2$ having equal control and $P_3$ having 0% effective control. Again, the basis for this conclusion is

---

[83] *Supra* note 61.
[84] *Id.*
[85] *Id.*

the number of times that the stockholders are critical voters in all possible coalitions. This is illustrated as follows:

| Possible Coalitions | Coalition Weight | Voting Outcome | Critical Stockholders | | |
|---|---|---|---|---|---|
| | | | $P_1$ | $P_2$ | $P_3$ |
| $\{P_1\}$ | 50 | Losing Coalition | | | |
| $\{P_2\}$ | 49 | Losing Coalition | | | |
| $\{P_3\}$ | 1 | Losing Coalition | | | |
| $\{P_1, P_2\}$ | 99 | Winning Coalition | | | |
| $\{P_2, P_3\}$ | 50 | Losing Coalition | | | |
| $\{P_1, P_3\}$ | 51 | Losing Coalition | | | |
| $\{P_1, P_2, P_3\}$ | 100 | Winning Coalition | | | |
| Number of Times that Stockholder is Critical | | | 2 | 2 | 0 |

The third example shows all stockholders having the same or equal degrees of control, which means that given a quota of 51%, the additional 10% voting weight of $P_1$ compared to the voting weights of $P_2$ and $P_3$ is immaterial in determining the outcome of a stockholder meeting. This is illustrated as follows:

| Possible Coalitions | Coalition Weight | Voting Outcome | Critical Stockholders | | |
|---|---|---|---|---|---|
| | | | $P_1$ | $P_2$ | $P_3$ |
| $\{P_1\}$ | 40 | Losing Coalition | | | |
| $\{P_2\}$ | 30 | Losing Coalition | | | |
| $\{P_3\}$ | 30 | Losing Coalition | | | |
| $\{P_1, P_2\}$ | 70 | Winning Coalition | | | |
| $\{P_2, P_3\}$ | 60 | Winning Coalition | | | |
| $\{P_1, P_3\}$ | 70 | Winning Coalition | | | |
| $\{P_1, P_2, P_3\}$ | 100 | Winning Coalition | | | |
| Number of Times that Stockholder is Critical | | | 2 | 2 | 2 |

The fourth example retains the same voting rights distribution as in the third example, but we changed the quota from a simple majority of 51% to a super-majority of 67%. With this change, the additional 10% voting weight of $P_1$ suddenly gains relevance, making him the stockholder with highest voting power. This is illustrated as follows:

| Possible Coalitions | Coalition Weight | Voting Outcome | Critical Stockholders | | |
|---|---|---|---|---|---|
| | | | $P_1$ | $P_2$ | $P_3$ |
| $\{P_1\}$ | 40 | Losing Coalition | | | |
| $\{P_2\}$ | 30 | Losing Coalition | | | |
| $\{P_3\}$ | 30 | Losing Coalition | | | |
| $\{P_1, P_2\}$ | 70 | Winning Coalition | | | |
| $\{P_2, P_3\}$ | 60 | Losing Coalition | | | |
| $\{P_1, P_3\}$ | 70 | Winning Coalition | | | |
| $\{P_1, P_2, P_3\}$ | 100 | Winning Coalition | | | |
| Number of Times that Stockholder is Critical | | | 3 | 1 | 1 |

## FORMAL DEFINITION OF VOTING POWER

We are now ready to provide a formal definition of voting power in a stockholder meeting. While the Control Test simply defines voting power as voting weight, we propose voting power as:

$$V_i = \frac{\beta_i}{\sum \beta_i}$$

$V_i$ denotes the voting power of a given stockholder $P_i$. $\beta_i$ denotes the number of times that stockholder $P_i$ is a critical stockholder in all possible stockholder coalitions. $\sum \beta_i$ denotes the total number of times that all stockholders are critical stockholders in all possible stockholder coalitions. [86] Applying this definition, we summarize the voting powers of stockholders in the four examples in the previous section, as follows:

| Voting Situations | $\beta_i$ | | | $\sum \beta_i$ |
| --- | --- | --- | --- | --- |
| | $P_1$ | $P_2$ | $P_3$ | |
| {51: 50, 49, 1} | 3 | 1 | 1 | 5 |
| {67: 50, 49, 1} | 2 | 2 | 0 | 4 |
| {51: 40, 30, 30} | 2 | 2 | 2 | 6 |
| {67: 40, 30, 30} | 3 | 1 | 1 | 5 |

The resulting voting power distribution is as follows:

| Voting Situations | $V_i$ | | |
| --- | --- | --- | --- |
| | $P_1$ | $P_2$ | $P_3$ |
| {51: 50, 49, 1} | 60% | 20% | 20% |
| {67: 50, 49, 1} | 50% | 50% | 0% |
| {51: 40, 30, 30} | 33.3% | 33.3% | 33.3% |
| {67: 40, 30, 30} | 60% | 20% | 20% |

Applying the formula for voting power, we reveal degrees of control that are not obvious when we merely look at the voting weight distributions of stockholders.

## *DE FACTO* FOREIGN CONTROL IN STOCKHOLDER MEETINGS

The thesis postulated in the introduction is that the Control Test does not guarantee that a foreign minority stockholder will have minority control. A stockholder does not have "minority" control if, empirically, he has equal or higher degree of control compared to other stockholders. Hence, there are two parameters that can falsify the Control Test: *first*, if the voting power of a foreign stockholder is equal to the voting power of each of the Filipino stockholders, and *second*, if his voting power is greater than that of each of the Filipino stockholders. Under the first parameter, we say that the foreign stockholder has "joint control", and under the second parameter, that he has *de facto* or "effective control".

---

[86] *Supra* note 6.

The question, therefore, is: when do foreign minority voting rights result in joint control, or *de facto* or effective control by foreigners? In other words, what are the instances when a corporation complies with a given foreign equity limitation, but a foreign minority stockholder has equal or more voting power compared to Filipino stockholders? Consider the following voting situations:

| Voting Weight Distribution ($P_1$ : foreigner) | Voting Power Distribution ($V_i$) | |
|---|---|---|
| | Simple Majority ($q = 51\%$) | Super-Majority ($q = 67\%$) |
| {60, 40} | {100%, 0%} | {50%, 50%} |
| {40, 60} | {0%, 100%} | {50%, 50%} |
| {40, 30, 30} | {33.33%, 33.33%, 33.33%} | {60%, 20%, 20%} |
| {40, 20, 20, 20} | {50%, 16.67%, 16.67%, 16.67%} | {40%, 20%, 20%, 20%} |
| {49, 51} | {0%, 100%} | {50%, 50%} |
| {49, 26, 25} | {33.33%, 33.33%, 33.33%} | {60%, 20%, 20%} |
| {49, 17, 17, 17} | {50%, 16.67%, 16.67%, 16.67%} | {40%, 20%, 20%, 20%} |
| {30, 70} | {0%, 100%} | {0%, 100%} |
| {30, 24, 23, 23} | {50%, 16.67%, 16.67%, 16.67%} | {25%, 25%, 25%, 25%} |
| {25, 75} | {0%, 100%} | {0%, 100%} |
| {25, 38, 37} | {33.33%, 33.33%, 33.33%} | {0%, 50%, 50%} |
| {25, 19, 19, 19, 18} | {20%, 20%, 20%, 20%, 20%} | {20%, 20%, 20%, 20%, 20%} |
| {20, 80} | {0%, 100%} | {0%, 100%} |
| {20, 40, 40} | {33.33%, 33.33%, 33.33%} | {0%, 50%, 50%} |
| {20, 27, 27, 26} | {0%, 33.33%, 33.33%, 33.33%} | {25%, 25%, 25%, 25%} |
| {20, 16, 16, 16, 16, 16} | {33.33%, 13.33%, 13.33%, 13.33%, 13.33%, 13.33%} | {30%, 14%, 14%, 14%, 14%, 14%} |

The table lists various voting weight distributions in corporations engaged in partially nationalized economic activities, with $P_1$ as the lone foreign stockholder. For every voting weight distribution, the voting weight of $P_1$ maximizes a given foreign equity limitation. Hence, in the {60, 40} distribution, foreign stockholder $P_1$ has a voting weight of 60%, which maximizes the allowable foreign equity in financing companies and investment houses regulated by SEC, as provided in Section 6 of R.A. 5980 as amended by R.A. 8556 and P.D. 129 as amended by R.A. 8366. In the {40, 60} distribution, foreign stockholder $P_1$ has a voting weight of 40%, which is also the maximum foreign equity in public utility companies, as provided in Section 11 of Article XII of the 1987 Constitution.

The voting weight distributions may pertain to voting shares, or to the total outstanding capital stock, which includes both voting and non-voting shares. As provided in *Gamboa vs. Teves*, the foreign equity cap applies to the total outstanding capital stock *and* to each class of shares, whether voting or non-voting.[87] Hence, the resulting voting power distribution of the given voting weight distribution is true whether the voting situation includes all or only some classes of shares. For instance, the voting weight distribution {51: 40, 30, 30}, which results in a {33.33%, 33.33%, 33.33%} voting power distribution, is true whether the context is a stockholder meeting requiring approval of submitted matters, which only involve common stockholders, or fundamental matters, which involve all stockholders, whether common or preferred shareholders.

---
[87] G.R. No. 176579, October 09, 2012

In all instances, $P_1$ is the lone foreigner in the corporation and all other stockholders are Filipinos. The different scenarios explore voting situations where there is only one, few or many Filipino stockholders, given the same voting weight for $P_1$. For instance, where $P_1$ maximizes a foreign equity cap of 40%, we explore scenarios where there is only one or two other Filipino stockholders. We also explore different combinations of voting weights between or among the Filipino stockholders. As shown in previous disquisitions, reconfiguring the voting weights of *other* stockholders can reconfigure the voting power of a stockholder whose voting weight remains unchanged.

The first column under voting weight distribution represents the *de jure* allocation of control in the corporation, which renders $P_1$ as a minority stockholder in terms of voting weight. The distribution merely reflects the shareholding size of each stockholder. Since the foreigner only occupies a *de jure* minority position[88], the corporation passes the Control Test, and is therefore considered a Philippine national for the purpose of complying with foreign equity limitations.

The second and third columns under voting power distribution represents the *de facto* allocation of control, which shows degrees of power that are not obvious if we only look at the voting weight distribution in the first column, as prescribed by the Control Test. For the voting power distribution, we apply the formal definition of voting power as $V_i$. This results in a *de facto* control allocation that is different from the *de jure* control allocation coming from the voting weight distribution.

We have explained in the previous sections how the quota can reconfigure the voting power of stockholders, even if the voting weight distribution remains unchanged. Hence, the voting power distribution is divided into two columns by quota: whether simple majority (51%) or super-majority (2/3 or 67%). We have excluded a column for a quota requiring unanimous votes (100%) since, naturally, this will render all the stockholders in all voting situations to have equal degrees of control, similar to a one person-one vote system.

Why should we give emphasis to simple majority and super-majority voting requirements? This is because the Corporation Code reserves certain matters to be decided by the stockholders and prescribes the corresponding quota. Apart from the selection of the members of the Board of Directors, matters requiring stockholder approvals include: (1) those required by law to be approved by the stockholders with voting shares; (2) those required by law to be approved by the stockholders regardless of whether they hold voting or non-voting shares; and (3) those submitted by management to the stockholders for approval, which by default only involve stockholders with voting shares, unless otherwise provided in the by-laws.[89]

The voting weight distribution and the quota allow us to derive $V_i$. The voting power distribution indicates that *it is possible for a Philippine national to be under the "effective control" of a foreign minority stockholder*. It is also possible to pass the Control Test, even though the foreign minority stockholder has "joint control" of the corporation.

---

[88] Except for the {60, 40} voting weight distribution, which is allowed by law.
[89] B.P. No. 68.

In the {60, 40} voting weight distribution and given a simple majority voting requirement, $P_1$ has "effective control" and is a dictator stockholder because his voting weight is higher than the quota. Therefore, his voting power is 100% and that of the Filipino stockholder, 0%, even though the latter has a 40% voting weight. It does not matter whether $P_2$ has 1% voting weight or 40%. By imposing a super-majority voting requirement, however, $P_1$ loses his dictator status because he can no longer unilaterally pass a stockholder resolution. Nevertheless, he has "joint control", and the Filipino stockholder ($P_2$) cannot pass a stockholder resolution without the cooperation of the foreign stockholder.

In the {40, 60} voting weight distribution, $P_1$ has virtually no control of the corporation, but in a matter requiring super-majority votes, $P_1$ has joint control. If we compare this to the {40, 30, 30} voting weight distribution, we see the instant effect of having more Filipino stockholders and of dispersing the 60% voting weight between them. With simple majority voting requirement, $P_1$ has joint control, but with super-majority voting requirement, $P_1$ has *de facto* or effective control. The effect of dispersing the Filipino bloc of shares to more Filipino stockholders is more obvious if we compare this further to the {40, 20, 20, 20} voting weight distribution, where $P_1$ has *de facto* control regardless of whether the vote calls for simple majority or super-majority.

We see the same pattern in the succeeding voting weight distributions, with foreign equity limitations of 49%, 30%, 25% and 20%. The voting power of foreign stockholder $P_1$ increases based on two factors: *first*, when the quota is raised, and *second*, when the total voting weight of Filipino stockholders is dispersed among an increasing number of Filipino stockholders. As we increase the quota, the number of Filipino stockholders, and the level of voting weight dispersion among them, we see an increasing progression in the voting power of the foreign stockholder. The voting power of $P_1$ changes from having no control to joint control, and finally from joint control to *de facto* or effective control. These changes occur even if the foreign stockholder's shareholding size remains unchanged and even if the corporation continues to be considered a Philippine national under the Control Test.

One question that arises is: why compare the voting power of a foreign stockholder with that of *each* of the Filipino stockholders, if the Control Test treats all Filipino stockholders as *one* coalition? This is because the Control Test does not account for the fact that a stockholder who deals with fewer stockholders to pass a resolution has higher voting power than a stockholder who needs to deal with more stockholders. The voting power distribution accounts for this fact. For example, in the {40, 20, 20, 20} voting weight distribution, the Control Test presumes as a matter of legal fiction that $P_2$, $P_3$ and $P_4$ will form a coalition of Filipino stockholders. This is the reason why the Control Test adds the voting weights of the individual Filipino stockholders to determine the Filipino coalition weight. However, in the voting power distribution, $V_i$ does not assume that stockholders of the same nationality will form a coalition. It measures all possible stockholder coalitions, and determines which stockholder has the most advantageous position based on the voting weight distribution and the quota.

We have discussed the possibility that a foreign minority stockholder has *de facto* control of stockholder approvals. The next section is equally important: the possibility of *de facto* foreign control of the Board of Directors, notwithstanding the fact that foreign stockholders can only elect their nominees to the Board to the extent of their foreign equity participation.

## *DE FACTO* FOREIGN CONTROL OF BOARD OF DIRECTORS

There are four types of voting situations in the corporation: (1) those involving common stockholders, (2) those involving all stockholders, (3) election by common stockholders of members of the Board of Directors, and (4) voting situations within the level of the Board of Directors.[90] The preceding section modeled the first two voting situations as a weighted voting game, involving players with voting weights and a quota.

The third voting situation cannot be modeled as a weighted voting game because the election of Directors does not involve two alternative motions (i.e. "yes" and "no"). Rather, it is a situation where the voting weight of a common stockholder is translated into his number of representatives in the Board of Directors. The common stockholder's number of representatives becomes a proxy for his voting weight in the Board of Directors. Therefore, the election of Board of Directors merely transposes the voting weight distribution of common stockholders into the voting weight distribution of nominees in the Board. But this is only true if we view the nominees of one stockholder as a single coalition.

The fourth voting situation, which only involves the level of the Board, is a one person-one vote system. The Directors do not vote as a coalition of nominees of their respective nominator stockholder. Each Director exercises his own discretion and is entitled to one vote. This characterizes the voting situation within the Board as a one person-one vote system, where each voter has equal degree of control as the others.

Notwithstanding the fact that each Director is entitled to only one vote and that each Director has equal degree of control as the others, we have postulated in a previous section (*The Stockholder Meeting as a Weighted Voting Game*) that we can reconfigure a one person-one vote system into a weighted voting game. To recall, we stated that the concept of voting weight becomes relevant if a group of Directors is conceived as a coalition, in which case the one person-one vote system becomes a weighted voting game from the perspective of the coalitions of Directors.

The Control Test, as applied in *Gamboa vs. Teves*, conceives the voting situation in the Board of Directors as a weighted voting game. It presumes, by legal fiction, that the number of nominees of a stockholder in the Board of Directors is a proxy of the stockholder's voting weight. Hence, the voting weight distribution of common stockholders determines the level of representation of each common stockholder in the Board of Directors, and therefore the voting weight of a *presumed* coalition of Director nominees mirrors the voting weight of a common stockholder. If so, the voting power distribution in the Board of Directors also mirrors the voting power distribution of common stockholders. This is illustrated as follows:

---

[90] B.P. No. 68.

| Voting Weight Distribution of Stockholders ($P_1$ : foreigner) | Board Representation in 10 Director Positions ($P_1$ : foreign nominees) | Voting Weight Distribution in the Board of Directors |
|---|---|---|
| {40, 60} | {4, 6} | {40, 60} |
| {40, 30, 30} | {4, 3, 3} | {40, 30, 30} |
| {40, 20, 20, 20} | {4, 2, 2, 2} | {40, 20, 20, 20} |

From the illustration, we see how the voting weight of a stockholder is transposed into the voting weight of his Director nominees in the Board of Directors, if his Director nominees are conceived as a single coalition in the Board. Thus, if $P_1$ has 40% voting weight in the corporation, this entitles him to four Director nominees in the Board, which is composed of 10 available seats. The four Directors, conceived as a unit, has a combined voting weight of 40% in the Board, equal to the voting weight of $P_1$ as a stockholder. From the voting weight distribution in the Board of Directors, we derive the voting power of each coalition of Director nominees, as follows:

| Voting Weight Distribution of Stockholders ($P_1$ : foreigner) | Board Representation ($P_1$ : foreign nominees) | Voting Weight Distribution of Board | Voting Power Distribution in Board of Directors ($V_i$) | |
|---|---|---|---|---|
| | | | Simple Majority ($q$ = 51%) | Super-Majority ($q$ = 67%) |
| {40, 60} | {4, 6} | {40, 60} | {0%, 100%} | {50%, 50%} |
| {40, 30, 30} | {4, 3, 3} | {40, 30, 30} | {33.33%, 33.33%, 33.33%} | {60%, 20%, 20%} |
| {40, 20, 20, 20} | {4, 2, 2, 2} | {40, 20, 20, 20} | {50%, 16.67%, 16.67%, 16.67%} | {40%, 20%, 20%, 20%} |

If the foreign equity limitation is 40%, and the foreign stockholder has maximized the cap, he can elect four nominees in the Board of Directors, which yields a voting weight of 40% in the Board. The voting power of the coalition of the foreign stockholder's director nominees, in turn, depends on the quota within the Board, the number of Filipino stockholders, the corresponding number of Director nominees of Filipino stockholders, and the voting weight of each Filipino stockholder. Thus, while a {40, 60} voting weight distribution results in 0% control in the Board for a simple majority voting requirement or joint control for a super-majority voting requirement, a {40, 30, 30} voting weight distribution results in a *de facto* control of the Board for a super-majority voting requirement, and a {40, 20, 20, 20} voting weight distribution results in a *de facto* control of the Board for both majority and super-majority voting requirements.

**EFFECT OF PUBLIC FLOAT**

The public float is that portion of a corporation's capital stock owned by an infinitely large number of stockholders.[91] This exists in publicly listed corporations.[92] Each holder of shares of stock in the public float, therefore, virtually possesses 0% degree of control, because each member in the public float must deal with an infinitely large number of stockholders to form a coalition of stockholders.[93] To recall, a stockholder who deals with fewer stockholders to pass a resolution has more power compared to a stockholder who needs to deal with more stockholders.[94] Hence, it is safe to assume that the public float will never vote as a single coalition, and their combined voting weight will not have an impact in the voting power of bloc-holders.[95]

---

[91] *Supra* note 6.
[92] *Id.*

A more realistic voting weight distribution of a corporation with public float must therefore exclude the combined voting weight of the public float.[96] For example, in a {40, 20, 20, 20} voting weight distribution, where $P_1$ is a foreign stockholder and $P_4$ represents the combined voting weight of the public float, we will not expect that an infinitely large number of stockholders will form a coalition to vote the combined voting weight of 20%.[97] Hence, we modify the voting weight distribution by deducting the public float's combined voting weight of 20% from the total outstanding shares of 100%. We then divide the remaining voting weights by the reduced amount of total outstanding shares. This yields a new voting weight distribution of {50, 25, 25}, *net* of public float. The corresponding changes in voting power are illustrated as follows:

|  | Voting Weight Distribution ($w_i$) | Voting Power Distribution ($V_i$) |
| --- | --- | --- |
| With Public Float | {51: 40, 20, 20, 20} | {50%, 16.67%, 16.67%, 16.67%} |
| Without Public Float | {51: 50, 25, 25} | {60%, 20%, 20%} |

Note that, with the reduction of the combined weight of the public float from the distribution, the voting power of $P_1$ increased by 10% while the individual voting powers of $P_2$ and $P_3$ increased only by 3.33%.

**EFFECT ON THE GRANDFATHER RULE**

The Grandfather Rule offers a deceptively simple formula for unraveling chains of control in a complex web of corporate layering. According to this rule, if Corporation A holds 60% shares in B, which holds 30% shares in C, it follows that A indirectly controls 18% of C (i.e. 60% multiplied by 30%).[98] This method of imputing a fractional share of indirect control, however, is based on two flawed assumptions. *First*, it assumes that voting situations across the chain are simultaneously occurring, when the reality is that stockholder votes happen sequentially from the first to the second tier of stockholders. *Second*, it treats voting power across the chain as a "continuous variable", when the more accurate method is to treat it as a "discrete variable".[99]

By imputing a fractional share of indirect control, it is as if the voting power of minority stockholders in Corporation B is still relevant in determining the result of a stockholder meeting in C once A already prevails in a given voting situation in B. The fact is that, with a simple majority voting requirement, A can single-handedly pass a motion in a stockholder meeting in B, since A's voting weight of 60% is already greater than the minimum threshold for approving the motion. The total 40% voting weight of other minority stockholders in B is immaterial in determining the outcome of the voting scenario. The more accurate method, therefore, is to treat voting power along the chain as a "discrete variable": *A is the controlling stockholder in B—true or false?*[100] And if the answer is true, then A is deemed to control the entirety of B's voting weight in C, and not just a fraction of 30%.

---

[93] *Id.*
[94] *Supra* note 61.
[95] Hart, S., *Shapley Value*, GAME THEORY, Palgrave Macmillan UK 210-216 (1989)
[96] *Supra* note 6.
[97] *Id.*
[98] See *Narra Nickel Mining and Development Corp. vs. Redmond Consolidated Mines Corp.*, G.R. No. 195580, January 28, 2015
[99] *Supra* note 6.

A variable is continuous if it can assume infinite values in an interval. This is true in the case of voting weights, where the possible values can be any real number between 0% and 100%. Voting power varies as the weights change in the continuum of infinite possible values. This variable type is appropriate in the Control Test, where a slight change in shareholding size can make a stockholder win or lose in a voting situation. For example, where stockholders D and E have weights of 50% each, a sudden shift of 1% from E to D can make D prevail in a voting scenario, even without the cooperation of E.

A variable is discrete if the possible values are countable and finite. If this is applied in measuring voting power across a chain of corporations, there are only two possible answers: whether the stockholder in question is dominant in the higher tier or not, using a pass-fail criterion. The determination of who is dominant, however, is more complex than what is provided by the Control Test.

To develop an alternative to the Grandfather Rule, we shall construe the question, "Who is the controlling stockholder?" as having two dimensions: horizontal and vertical.[101] Horizontal control refers to the voting power exercised by a stockholder relative to other stockholders in a corporation.[102] Vertical control refers to the voting power exercised by a stockholder across a chain of corporations.[103] In the law on foreign investments, the Control Test measures horizontal control, while the Grandfather Rule measures vertical control. Horizontal control is continuous, while vertical control should be discrete.

Regardless of whether we measure horizontal control through voting weight or voting power, both of them are continuous variables. The same variable type should not be applied for measuring vertical control. Voting situations happen successively or consecutively, i.e. from the highest tier to the lowest tier of stockholder corporations. The proper approach, therefore, is to determine the controlling stockholder in every tier, and impute a non-fractional indirect control to the stockholder. To illustrate, consider the following ownership structure (*Figure 1*):

---

[100] *Id.*
[101] *Id.*
[102] *Id.*
[103] *Id.*

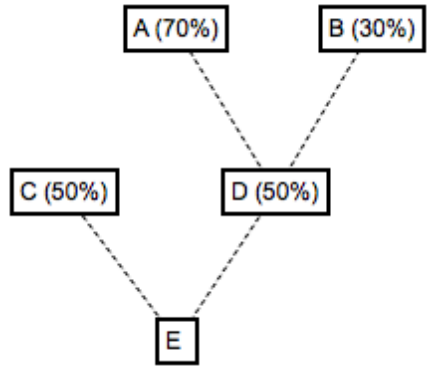

Under the Grandfather Rule, Corporations A and B indirectly control E through D. To impute the indirect shareholding, the 70% equity of A and 30% equity of B are multiplied with D's 50% equity in E, in order to arrive at a hypothetical shareholding structure with no indirect holding (*Figure 2*), as follows:

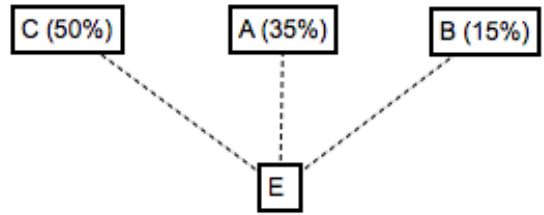

It is a fallacy to split the 50% equity holding in E between A and B, according to the proportion of their equity holding in D. *First*, this is a voting scenario that will not happen in reality because voting across tiers occur in a time series: first between A and B, and then between C and D-as-controlled-by-A. *Second*, the voting power distribution in the last tier of stockholders in Figure 1 is very different from the voting power distribution in Figure 2. In Figure 1, A can unilaterally pass a stockholder resolution and B's voting weight is immaterial in determining the voting scenario. Hence, when we reach the voting scenario in the second tier, C's voting power is co-equal with D-as-controlled-by-A, which means that C can veto D's motion, and D can veto C's motion. In Figure 2, however, we see the voting power of C dilated and that of A diluted by the presence of all three stockholders in the tier and by the fractional share of indirect control. The possible winning coalitions are: {C, A}, {C, B} and {C, A, B}, but not {A, B}. Based on this, A has no veto power against any motion, when in reality A controls the entire voting weight of D and is therefore entitled to veto a motion if A so desires. In other words, the Grandfather Rule understates the indirect control held by A, as follows:

| Control Measurements | Figure 1 | Figure 2 |
|---|---|---|
| Voting Weight | C (50%), D (50%) | C (50%), A (35%), B (15%) |
| Voting Power | C (50%), D (50%) | C (60%), A (20%), B (20%) |

Without the Grandfather Rule (*Figure 1*), the voting weight in the last tier of stockholders reflects their voting power. After applying the Grandfather Rule (*Figure 2*), we see a discrepancy of voting weight and voting power. The increase in the effective control of C from a voting weight of 50% to a voting power of 60% reflects the fact that C is absolutely needed in every possible stockholder coalition to pass a motion, while the decrease in the effective control of A from a voting weight of 35% to a voting power of 20% reflects the fact that A is not indispensable to pass a stockholder resolution. It is therefore meaningless to say that A only has 35% indirect control under the Grandfather Rule. The more realistic description of the chain of control is as follows:

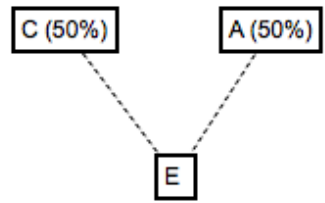

Revisiting *Narra Nickel* and *Gamboa*

What is the effect of the voting power computation on the Supreme Court rulings in *Narra Nickel Mining vs. Redmont Consolidated Mines Corp.* and *Gamboa vs. Teves*? Applying the methodology discussed in the previous sections, our findings indicate that in *Narra Nickel*, Filipino stockholders have a degree of control equal to that of the foreign stockholder under a super-majority setup, and effective control under a simple majority setup, contrary to the ruling of the Supreme Court which accords effective control to the foreign stockholder. In *Gamboa*, our findings indicate that, as of March 2016, the foreign stockholder has a degree of control equal to that of the Filipino stockholders and therefore higher than what the 1987 Constitution allows, both under a simple majority and super-majority setup, notwithstanding the fact that the PLDT shareholding structure may be compliant within the framework of the *Gamboa* ruling.

**A. *NARRA NICKEL MINING VS. REDMONT CONSOLIDATED MINES CORP.***

In order to undertake exploration and mining activities, a corporation must apply for a Mineral Production Sharing Agreement (MPSA) and Exploration Permit (EP) with the Department of Environment and Natural Resources (DENR). Sara Marie Mining, Inc. (SMMI) applied for an MPSA and EP covering certain areas in the Province of Palawan. SMMI assigned its rights under the MPSA application to Madridejos Mining Corporation (MMC), and MMC further assigned them to McArthur Mining, Inc. ("McArthur"). Subsequently, SMMI again applied for another MPSA covering another area of Palawan. SMMI assigned its rights under the second MPSA application to Tesoro Mining and Development, Inc. ("Tesoro"). On a separate occasion, Alpha Resources and Development Corporation (ARDC) and Patricia Louise Mining & Development Corporation (PLMDC) applied for an MPSA in other areas of Palawan. PLMDC, in turn, assigned its rights under the MPSA application to Narra Nickel Mining and Development Corp. ("Narra Nickel").

McArthur, Tesoro and Narra Nickel were the existing right-holders under the MPSA applications when Redmont Consolidated Mines Corp. ("Redmont") took interest in undertaking exploration and mining activities in areas of Palawan already covered by the said MPSA applications. Redmont filed petitions for denial with the Panel of Arbitrators (POA) under the DENR, alleging that MBMI Resources, Inc. ("MBMI") owned at least 60% of the capital stock of McArthur, Tesoro and Narra Nickel. This would disqualify the MPSA applicants from undertaking mining and exploration activities in Palawan for being foreign nationals, pursuant to Section 3(aq) of Republic Act No. 7492 (*Philippine Mining Act of 1995*), which states that a "Qualified person" which is also a corporation must have at least 60% of the capital owned by citizens of the Philippines. MBMI is a 100% Canadian corporation.

Moreover, Section 2, Article XII of the 1987 Constitution provides, "The exploration, development, and utilization of natural resources shall be under the full control and supervision of the State. The State may directly undertake such activities, or it may enter into co-production, joint venture or production-sharing agreements with Filipino citizens, or corporations or associations at least sixty per centum of whose capital is owned by such citizens."

The corporate ownership structures of McArthur, Tesoro and Narra Nickel are illustrated as follows:

### Corporate Ownership Structure of McArthur

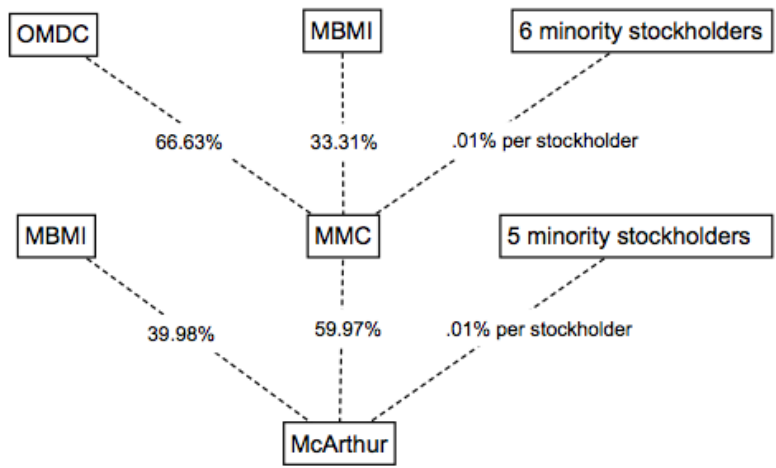

### Corporate Ownership Structure of Tesoro

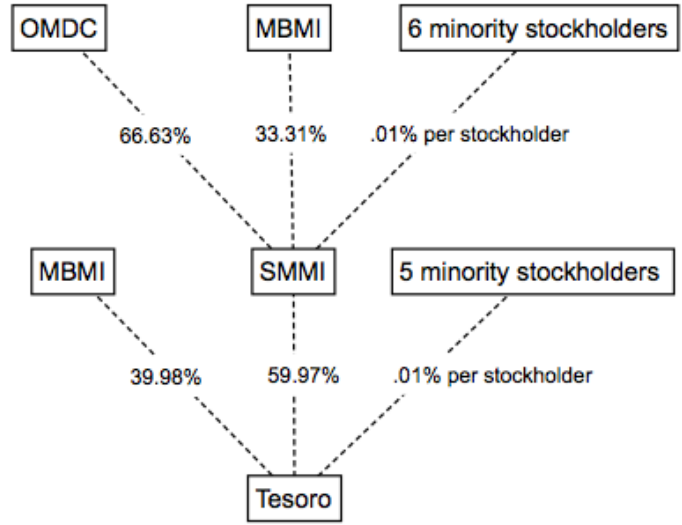

**Corporate Ownership Structure of Narra Nickel**

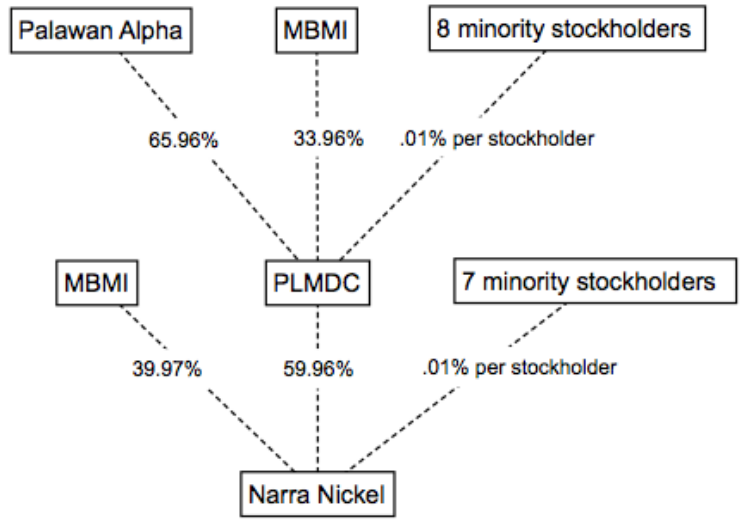

      In McArthur, MBMI is a Canadian corporation, OMDC is a Filipino corporation, and MMC is a Filipino corporation. At issue is the corporate nationality of McArthur. Since MMC is partly owned by a Filipino corporation (OMDC) and a Canadian corporation (MBMI), the Supreme Court bifurcated the 59.97% stockholding of MMC in McArthur as consisting partly of Filipino equity and partly of Canadian equity. Hence, of the 59.97% equity of MMC, 39.96% (66.63% x 59.97%) represents indirect Filipino equity while 19.98% (33.31% x 59.97%) represents indirect Canadian equity. The 19.98% indirect Canadian equity through MMC's stockholding is then added to 39.98% direct Canadian equity represented by the direct shareholding of MBMI in McArthur, resulting in an effective total Canadian equity in McArthur of 59.96%, which exceeds the 40% foreign equity limitation for mining and exploration activities. The Filipino equity in McArthur consists of the indirect equity from OMDC, which is 39.96%, and the negligible

equity from Filipino minority stockholders.

In Tesoro, MBMI has a direct Canadian equity of 39.98%, added to an indirect Canadian equity through SMMI of 19.98% (33.31% x 59.97%), for an effective total Canadian equity of 59.96%. In Narra Nickel, MBMI has a direct Canadian equity of 39.97%, added to an indirect Canadian equity through PLMDC of 20.36% (33.96% x 59.96%), for an effective total Canadian equity of 60.33%.

It is true that, with respect to dividend rights and other economic rights in McArthur, Tesoro and Narra Nickel, the Canadian national has effective total economic rights exceeding the 40% foreign equity limitation once we add the direct and indirect shareholdings of MBMI. But the same cannot be said about corporate control. Applying the methodology for computing voting power discussed in the previous sections, it is erroneous to conclude that MBMI has an effective voting power of 59.96% in McArthur, 59.96% in Tesoro, and 60.33% in Narra Nickel.

Each of the three corporate ownership structures have two tiers. To determine the corporate nationality of McArthur based on control, the first step is to determine the voting power distribution in the upper tier and identify the stockholder with effective control. The second step is to impute the entirety of the indirect shareholding to the controlling stockholder identified in the first step. The third step is to determine the voting power distribution in the lower tier and identify the stockholder with effective control. For McArthur, we analyze the upper tier as follows:

| Stockholders of MMC | Voting Weight Distribution in MMC | Voting Power Distribution in MMC ($V_i$) | |
| --- | --- | --- | --- |
| | | Simple Majority ($q = 51\%$) | Super-Majority ($q = 67\%$) |
| {OMDC, MBMI, 6 minority stockholders} | {66.63%, 33.31%, .01% per stockholder} | {100%, 0%, 0%} | {100%, 0%, 0%} |

The voting weights of the Filipino corporation (OMDC) and the Canadian corporation (MBMI) in MMC are 66.63% and 33.31%, respectively. But regardless of whether the threshold for passing stockholder resolutions is based on simple majority or super-majority (2/3), OMDC is a "dictator stockholder", i.e. capable of passing stockholder resolutions without the cooperation of other stockholders. The substantial stockholding of 33.31%, which represents Canadian equity, has an effective control equivalent to 0%. Hence, for the purpose of analyzing the voting power distribution in the lower tier, we must impute the entire stockholding of MMC in McArthur as belonging solely to OMDC. Analysis of the lower tier is as follows:

| Stockholders of McArthur | Voting Weight Distribution in McArthur | Voting Power Distribution in McArthur ($V_i$) | |
| --- | --- | --- | --- |
| | | Simple Majority ($q = 51\%$) | Super-Majority ($q = 67\%$) |
| {MBMI, MMC, 5 minority stockholders} | {39.98%, 59.97%, .01% per stockholder} | {0%, 100%, 0%} | {50%, 50%, 0%} |

The voting power distribution in the lower tier is one where the decision threshold becomes material. If the threshold is simple majority, MMC (as effectively controlled by OMDC) is a "dictator stockholder". However, if the threshold is 2/3 or super-majority, MMC and MBMI have joint control.

Considering the similarity in the three corporate ownership structures of McArthur and Tesoro, we arrive the same conclusions in Tesoro. The ownership structure of Narra Nickel, however, is slightly different. Analysis of the upper tier is as follows:

| Stockholders of PLMDC | Voting Weight Distribution in PLMDC | Voting Power Distribution in PLMDC ($V_i$) | |
| --- | --- | --- | --- |
| | | Simple Majority ($q = 51\%$) | Super-Majority ($q = 67\%$) |
| {Palawan Alpha, MBMI, 8 minority stockholders} | {65.96%, 33.96%, .01% per stockholder} | {100%, 0%, 0%} | {50%, 50%, 0%} |

The difference is that under a super-majority (2/3), the Filipino corporation (Palawan Alpha) and the Canadian corporation (MBMI) have joint or equal control of PLMDC. Analysis of the lower tier is as follows:

| Stockholders of Narra Nickel | Voting Weight Distribution in Narra Nickel | Voting Power Distribution in Narra Nickel ($V_i$) | |
| --- | --- | --- | --- |
| | | Simple Majority ($q = 51\%$) | Super-Majority ($q = 67\%$) |
| {MBMI, PLMDC, 7 minority stockholders} | {59.96%, 39.97%, .01% per stockholder} | {100%, 0%, 0%} | {50%, 50%, 0%} |

Once more, under a super-majority (2/3), PLMDC and MBMI have joint or equal control. While this may represent veto power, it neither represents minority control nor effective control. Whether the decision threshold is simple majority or super-majority, the conclusion we arrive at is different from the ruling of the Supreme Court, which imputes effective control of McArthur, Tesoro and Narra Nickel to the foreign national. Our findings indicate that the Filipino corporation has effective control under a simple majority setup, and has at least equal control under a super-majority setup.

### B. *GAMBOA VS. TEVES*

PLDT has an existing franchise to operate a telecommunications business in the Philippines. Based on the 2010 General Information Sheet (GIS) of PLDT, foreigners hold 120,046,690 common shares of PLDT while Filipinos hold 66,750,622 common shares; hence, foreign stockholders have 64.27% voting weight while Filipino stockholders have 35.73%. The Supreme Court ruled that "[s]ince holding a majority of the common shares equates to control, it is clear that foreigners exercise control over PLDT. Such amount of control unmistakably exceeds the allowable 40 percent limit on foreign ownership of public utilities expressly mandated in Section 11, Article XII of the Constitution."

As of March 2016, the shareholding structure of PLDT insofar as common shares are concerned is as follows:

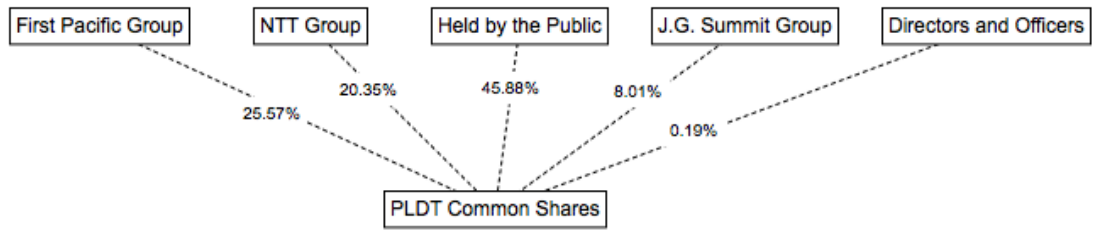

The voting power analysis yields the following initial results:

| Common stockholders of PLDT | Voting Weight Distribution among PLDT common stockholders | Voting Power Distribution among PLDT common stockholders ($V_i$) | |
| --- | --- | --- | --- |
| | | Simple Majority ($q = 51\%$) | Super-Majority ($q = 67\%$) |
| {First Pacific Group, NTT Group, Held by the Public, J.G. Summit Group, Directors and Officers} | {25.57%, 20.35%, 45.88%, 8.01%, 0.19%} | {16.67%, 16.67%, 50%, 16.67%%, 0%} | {30%, 10%, 50%, 10%, 0%} |

Considering, however, that "Held by the Public" shares and shares held by "Directors and Officers" constitute a large number of individual stockholders, with each having separate but negligible voting rights in PLDT, it is erroneous to treat them as blockholders capable of voting their shares as a single unit. Accordingly, we must apply the rules discussed under the section entitled Effect of Public Float to factor out the shareholdings represented by dispersed shareholders. For purposes of computing voting power, the modified shareholding structure only includes the blockholders First Pacific Group, NTT Group and J.G. Summit Group, with the following modified voting weights: {47.41%, 37.73%, 14.85%}. Analysis of this voting weight distribution is as follows:

| Common stockholders of PLDT | Voting Weight Distribution among PLDT common stockholders | Voting Power Distribution among PLDT common stockholders ($V_i$) | |
| --- | --- | --- | --- |
| | | Simple Majority ($q = 51\%$) | Super-Majority ($q = 67\%$) |
| {First Pacific Group, NTT Group, J.G. Summit Group} | {47.41%, 37.73%, 14.85%} | {33.33%, 33.33%, 33.33%} | {50%, 50%, 0%} |

Under a simple majority setup, NTT Group (which represents foreign equity) has joint or equal control as the other Filipino stockholders. Under a super-majority setup, NTT Group has power to veto the motions of First Pacific Group. This accords a degree of *de facto* control to the foreign stockholder higher than what the 1987 Constitution allows.

The intent to bestow veto power to the NTT Group is evident in its 2011 Annual Report, as duly filed with the U.S. Securities and Exchange Commission pursuant to Section 13 or 15(d) of the Securities Exchange Act of 1934, and filings with the Philippine Stock Exchange in 2012. PLDT discloses the content of a Shareholders Agreement bestowing "contractual veto rights" to the NTT Group, as follows:

> As a result of the Cooperation Agreement, NTT Communications and NTT DOCOMO, in coordination with each other, have contractual veto rights over a number of major decisions and transactions that PLDT could make or enter into, including:
>
> - capital expenditures in excess of US$50 million;
>
> - any investments, if the aggregate amount of all investments for the previous 12 months is greater than US$25 million in the case of all investments to any existing investees and US$100 million in the case of all investments to any new or existing investees, determined on a rolling monthly basis;
>
> - any investments in a specific investee, if the cumulative value of all investments made by us in that investee is greater than US$10 million in the case of an existing investee and US$50 million in the case of a new investee;
>
> - issuance of common stock or stock that is convertible into common stock;
>
> - new business activities other than those we currently engage in; and
>
> - merger or consolidation.

## CONCLUSION: THREE PRINCIPLES OF A RIGOROUS CONTROL TEST

Based on the foregoing ratiocinations, we postulate three principles of a rigorous test of voting power, which are lacking in the Control Test: (1) monotonicity, (2) *a prioricity*, and (3) probability. "Monotonicity" means that as the value of the voting power measurement increases or decreases, the actual degree of control that it describes likewise increases or decreases. "*a prioricity*" means that, in measuring degrees of control, we make no assumptions about the preferences of stockholders in forming coalitions with other stockholders. "Probability" is a consequence of *a prioricity*: since we make no assumptions about stockholder preferences, the voting power measurement must consider all possible stockholder coalitions and their corresponding voting outcomes, in order to determine which stockholder is most or least likely to dictate a voting scenario. We explain further as follows:

1. *Monotonicity*. – The higher is the amount of shares of stock owned by a stockholder, the higher is the amount of economic rights. This is not true in the case of control or voting rights, as measured by the Control Test. As demonstrated in the previous sections, increasing voting rights does not necessarily increase voting power, and decreasing voting rights does not necessarily decrease voting power. Hence, shareholding size or voting weight has a "non-monotonic" relationship with voting power. Contrast this with the voting power measurement we discussed in the previous sections, where a higher $V_i$ indicates more control and a lower figure indicates lesser control. In this method of measuring control, the magnitude of a stockholder's $V_i$ is "monotonic" with the actual degree of control in the corporation.

2. *a prioricity*. – The Control Test, as applied by the courts in corporate nationality disputes, makes *a priori* assumptions about the preferences of stockholders in forming coalitions. For example, given a voting weight distribution of {40, 20, 20, 20} with $P_1$ as foreign stockholder, and given a foreign equity cap of 40%, the Control Test adds up the individual voting weights of $P_2$, $P_3$ and $P_4$, for a combined weight of 60%. The Control Test therefore assumes a fictional voting scenario where the three Filipino stockholders will combine to form a coalition in order to block a motion by $P_1$. It is because of this assumption that the Control Test treats the corporation as a Philippine national, since it is under the *de jure* control of Filipino stockholders.

    In our proposal, we do not make an assumption about stockholder preferences. Absent any prior information, the *a priori* assumption should be that all possible stockholder preferences are equally likely. In the {40, 20, 20, 20} distribution, there is no basis to combine the voting weights of Filipino stockholders by virtue of their common nationality. $P_1$ is just as likely to form a coalition with $P_2$ as $P_2$ is likely to form a coalition with $P_3$ or $P_4$.

3. Probability. – Considering that we have no *a priori* information about stockholder preferences in forming coalitions, a rigorous Control Test must consider all possible stockholder preferences and therefore, all possible coalitions. Each possible coalition contains information about which stockholder can make the coalition win or lose in a given voting scenario. The formula for $V_i$ measures the frequency of this information, from which is derived the likelihood that a stockholder's motion will prevail.

This essay lays down the theoretical foundation of a rigorous voting power measurement, as applied in the law on the determination of corporation nationality for the purpose of foreign investments. As agenda for future research, the logical next step is to survey the current state of voting power structures in corporations engaged in partially nationalized economic activities, with the aim of ascertaining whether—using voting power index in cooperative game theory—Filipino majority stockholders truly have "effective control" of said corporations given their existing ownership structures.